\documentclass[conference]{IEEEtran}
\IEEEoverridecommandlockouts
\usepackage{cite}
\usepackage{amsmath,amssymb,amsfonts}
\usepackage{algorithmic}
\usepackage{graphicx}
\usepackage{textcomp}
\usepackage{xcolor}
\def\BibTeX{{\rm B\kern-.05em{\sc i\kern-.025em b}\kern-.08em
    T\kern-.1667em\lower.7ex\hbox{E}\kern-.125emX}}
\begin{document}

\title{An Audio-Based Deep Learning Framework For BBC Television Programme Classification
}
\author{Lam~Pham$^{1,2}$, 
             Chris~Baume$^{3}$, 
             Qiuqiang~Kong$^{4}$,
             Tassadaq Hussain$^{2}$,
             Wenwu~Wang$^{2}$, 
             Mark~Plumbley$^{2}$ \\ \\
1. Center for Digital Safety \& Security, Austrian Institute of Technology, Austria.\\
2. Center for Vision Speech and Signal Processing, University of Surrey, UK \\
3. BBC Research and Development, BBC, UK \\
4. ByteDance AI Lab, ByteDance, US \\
}

\maketitle

\begin{abstract}
This paper proposes a deep learning framework for classification of BBC television programmes using audio.
The audio is firstly transformed into spectrograms, which are fed into a pre-trained Convolutional Neural Network (CNN), obtaining predicted probabilities of sound events occurring in the audio recording. 
Statistics for the predicted probabilities and detected sound events are then calculated to extract discriminative features representing the television programmes.
Finally, the embedded features extracted are fed into a classifier for classifying the programmes into different genres.
Our experiments are conducted over a dataset of 6,160 programmes belonging to nine genres labelled by the BBC. 
We achieve an average classification accuracy of 93.7\% over 14-fold cross validation. This demonstrates the efficacy of the proposed framework for the task of audio-based classification of television programmes.
\end{abstract}


\section{Introduction}
As the most popular media source in the UK, the BBC is used by more than 90\% of adults every week~\cite{bbc_report}. 
We wish to develop an effective recommendation system that helps audiences find suitable programmes based on their interests and needs.
Achieving an effective recommendation system not only requires a diverse user profile, but detailed metadata about the content.
However, this is challenging for broadcasters as their content is neither segmented nor well-defined.
Creating metadata manually is expensive, so an automatic tool to extract relevant metadata describing the content of BBC programmes reduces the cost.
Given a BBC programme, diverse resources such as topic, video (image data), transcript (text data), or audio (acoustic data), which contain rich information, can be used for extracting metadata.
We focus on analysing information extracted from audio, to evaluate whether both natural sounds and human speech detected in BBC programmes are useful for generating programme metadata.
To evaluate the value of audio, we firstly propose a task of audio-based classification of television programmes in this paper which makes use of deep learning techniques. 
In particular, the task proposed is to classify audio recordings of BBC programmes into the nine BBC genre categories~\cite{genres}: \textit{Children's, Drama, Factual, Music, Sport, Weather, Comedy, Entertainment} and \textit{News}. 
Given the deep learning classification model achieved in this paper, audio feature will be extracted and then integrated into the BBC metadata.
Additionally, the effect of audio feature on the BCC recommendation system will be evaluated and compared with the topic-based, text-based, or image-based features \footnote{The task of evaluating the role of audio features over the BBC recommendation system is our future work that is not presented in this paper.}.

\section{Background}
Regarding recently proposed systems for broadcast media classification~\cite{task_00, task_01, task_02, task_03, task_04, task_08}, authors made use of multiple features extracted from various resources such as audio, video, or transcript to maximize the system performance. 
For example, systems proposed in~\cite{task_03, task_04} explored the programme context via semantic concepts of sunset, indoor, outdoor, cityscape, landscape, mountains, etc., and their relation among classified genres.
Meanwhile, both audio and video features were made use in~\cite{task_00, task_01, task_02, task_08}.
To further improve a real-time system performance, an Automatic Speech Recognition (ASR) model was used in~\cite{task_05} to detect key words that enriches the word vectors representing a programme.
Recently, Mortaza \textit{et al.}~\cite{task_07_01} provided a comprehensive analysis of main features such as audio, text and the other metadata (channel, time) that were used for BBC broadcast classification.
%
\begin{figure*}[t]
    \centering
    \includegraphics[width =1.0\linewidth]{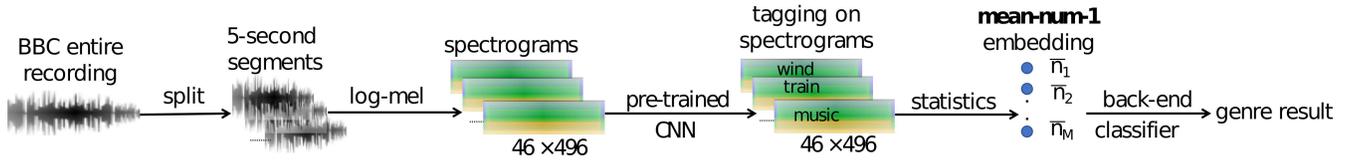}
	\caption{Baseline architecture}
    \label{fig:A1}
\end{figure*}
From the existing literature of audio feature extraction in broadcast classification ~\cite{task_08, task_06_01, task_07_01, task_10}, we can see that these methods followed two main steps: Acoustic modeling on short-time segments and statistical modeling across the programme. 
For example, Ekenel \textit{et al.}~\cite{task_08} firstly extracted Mel-frequency cepstral coefficients, fundamental frequency, signal energy, zero crossing rate from short-time segments split from each programme. Then, they applied a Gaussian Mixture Model (GMM) to learn these features, resulting in an embedded vector containing mean, standard deviation and weight of the GMM models.
Similarly, in~\cite{task_06_01, task_07_01, task_10}, short-time segments split from each audio recording were firstly transformed into vectors by using Linde--Buzo--Gray vector quantization algorithm~\cite{ref_01}. These quantized vectors were then trained by a GMM model, resulting in embedded vectors containing GMM model information. Finally, latent Dirichlet allocation (LDA) based statistics, which have been used mostly in natural language processing (NLP) for the categorisation of text documents, were applied to map embedded vectors to the LDA domain.
These audio feature extraction methods show how GMM was widely used for acoustic modeling over short-time segments.
However, short-time segments in each television programme may contain very different or very similar sound events. For example, a short-time segment may contain different natural sound events such as wind, bird song, or water fall sounds in \textit{Factual} programmes, or only content human speech in \textit{News} or \textit{Weather} programmes.
Therefore, using GMM for acoustic modeling over short-time segments may be ineffective with BBC programmes.
Indeed, a performance comparison presented in~\cite{task_07_01} indicates that audio feature obtained low performance compared to the other features.
Additionally, some television programmes have a long duration -- often more than four hours for events such as live sports. This leads to a high cost for training GMM models.
To deal with the limitations of GMM for acoustic modelling within the proposed task, a deep learning based framework is proposed.
In particular, we firstly train a CNN based network with AudioSet dataset~\cite{audioset} defining $M$ types of natural sounds and human speeches in life. 
Short-time segments split from each programme recording are transformed into spectrograms, then fed into the pre-trained CNN network to obtain the predicted probabilities for each of the $M$ sound events defined in AudioSet dataset. 
We then conduct statistics over the predicted probabilities and sound events detected on all segments to generate embedded features representing each television programme. 
Given by the embedding extracted, we classify it into one of the nine different genres mentioned earlier.
\begin{table}[t]
    \caption{The pre-trained CNN based network architecture (input image patch of $64{\times}496$)} 
        	\vspace{-0.2cm}
    \centering
    \scalebox{0.95}{
    \begin{tabular}{l c} 
        \hline 
            \textbf{Network architecture}   &  \textbf{Output}  \\
        \hline 
         Convolution [$3{\times}3 $@$ 64$] - BN - ReLU  & $64{\times}496{\times}64$\\
         Convolution [$3{\times}3 $@$ 64$] - BN - ReLU - Dropout (20\%)& $64{\times}496{\times}64$\\
         
         Convolution [$3{\times}3 $@$ 128$] - BN - ReLU  & $64{\times}496{\times}128$\\
         Convolution [$3{\times}3 $@$ 128$] - BN - ReLU - Dropout (20\%)  & $64{\times}496{\times}128$\\
         
         Convolution [$3{\times}3 $@$ 256$] - BN - ReLU  & $64{\times}496{\times}256$\\
         Convolution [$3{\times}3 $@$ 256$] - BN - ReLU - Dropout (20\%)  & $64{\times}496{\times}256$\\
         
         Convolution [$3{\times}3 $@$ 512$] - BN - ReLU  & $64{\times}496{\times}512$\\
         Convolution [$3{\times}3 $@$ 512$] - BN - ReLU  & $64{\times}496{\times}512$\\      
         
         Average Pooling [$2\times2$] - Dropout (30\%)& $32{\times}248{\times}512$\\         
         
         Convolution [$3{\times}3 $@$ 1024$] - BN - ReLU  & $32{\times}248{\times}1024$\\
         Convolution [$3{\times}3 $@$ 1024$] - BN - ReLU - Dropout (30\%)  & $32{\times}248{\times}1024$\\
         
         Convolution [$3{\times}3 $@$ 2048$] - BN - ReLU  & $32{\times}248{\times}2048$\\
         Convolution [$3{\times}3 $@$ 2048$] - BN - ReLU  & $32{\times}248{\times}2048$\\         
         
         Global Pooling - Dropout (50\%)  &  $2048$ \\
         FC - ReLU - Dropout (50\%)&  $2048$       \\
         FC - Softmax  &  $M=527$       \\
       \hline 
    \end{tabular}
    }
    \label{table:CDNN} 
\end{table}

\section{Baseline Architecture}
\label{baseline}

To evaluate the framework proposed, we firstly design a baseline system, outlined in Fig. \ref{fig:A1}.
As shown in Fig. \ref{fig:A1}, the baseline proposed comprises four main steps: the spectrogram transformation,  the pre-trained CNN--based network, statistics for embedded feature extraction, and the back-end classifier. Each is described below.

\subsection{Log-mel Spectrogram Transformation}
As the duration of BBC programmes can be very long, the programme recording is firstly split into non-overlapping 5-second segments which are suitable for back-end classifiers. 
The 5-second segments are then transformed into log-mel spectrograms of $46\times496$ by using Librosa toolbox~\cite{librosa_tool}. We re-used settings from our previous work~\cite{kong_pretrain}: fMin = 50 Hz, fMax = 14000 Hz, sample rate = 32000 Hz, window size = 1024, hop size = 320, and Mel filter number = 64.

\subsection{Pre-trained CNN--based Network}
\label{emb}

As a part of our previous work~\cite{kong_pretrain}, the pre-trained CNN based network, which was trained with AudioSet dataset~\cite{audioset}, is based on the VGG architecture~\cite{vgg_net} as shown in Table \ref{table:CDNN}.
In particular, the architecture contains sub-blocks which perform convolution, batch normalization (BN)~\cite{batchnorm}, rectified linear units (ReLU)~\cite{relu}, average pooling, global pooling~\footnote{the global pooling used is a combination of both max and average global pooling~\cite{kong_pretrain}}, dropout~\cite{dropout}, fully-connected (FC) and Softmax layers. 
The dimension of Softmax layer is set to $M=527$ that equals to the number of sound events defined in AudioSet dataset~\cite{audioset}.
In total, we have 12 convolutional layers and two fully-connected layers containing trainable parameters that makes the proposed CNN network like VGG-14~\cite{vgg_net}. 

\subsection{Statistics For Embedded Feature Extraction}
Given the pre-trained CNN network architecture, when a 5-second log-mel spectrogram is fed into the network, the output of the Softmax layer, likely a $M$-dimensional vector, is obtained.
Each dimension of the vector presents the predicted probability of one type of $M$ sound events that may occur in each 5-second segment.
In the baseline system proposed, we only select the sound event which shows the highest probability for extracting the embedded feature. 
In other words, each segment in a programme is now tagged by only one sound event with the highest probability, referred to as single-sound-event tagging information. 
Let us consider $\mathbf{n} = (n_{1}, n_{2},\ldots,n_{M})$ as the \textit{number} vector, where $M$ is the number of sound events defined in AudioSet dataset~\cite{audioset}, and $n_{i}$ is the total number of times that the $i$th sound event is detected and used to tag on segments in each programme.
The embedded feature $\mathbf{mean{-}num{-}1}\footnote{Note that the number `$1$' in the embedding name is used to reflect that only one sound event is used to tag on one segment.} = (\bar{n}_{1}, \bar{n}_{2},\ldots,\bar{n}_{M})$ is computed by $\bar{n}_{i} = \frac{n_{i}}{\sum_{i=1}^{M}n_{i}}$.
 %
%
\subsection{Back-end Classifier}
Given the embedding $\mathbf{mean{-}num{-}1}$, the baseline system uses Linear Regression for classifying them into nine genres of \textit{Children's, Drama, Factual, Music, Sport, Weather, Comedy, Entertainment}, and \textit{News}. 

\section{Candidate Architectures}

\subsection{Candidate Statistics For Embedded Feature Extraction}

With the baseline system proposed in Section \ref{baseline}, only one sound event with the highest probability is used to tag on each 5-second audio segment. 
This may lead to a reduction in system performance in cases that certain sound events are dominant in a television programme. 
For example, as human speech is dominant in \textit{News} and \textit{Weather}, it is easy to misclassify between these two genres.
This inspires us to evaluate whether multiple-sound-event tagging information (i.e. one segment is tagged by multiple sound events) is beneficial for representing each 5-second segment.
Therefore, we propose to use $k$ ($k = \{4, 6, 8, 10\}$) sound events which occupy top-$k$ predicted probabilities to tag on each 5-second segment.
We thus call $\mathbf{mean{-}num{-}k}$ as the embedded feature extracted when using the top-$k$ sound events for tagging.
Compute the embedding $\mathbf{mean{-}num{-}k}$
is same as $\mathbf{mean{-}num{-}1}$ proposed in the baseline system.

Furthermore, we evaluate whether predicted probabilities are beneficial for extracting embedded features.
We refer to $\mathbf{mean{-}prob{-}k}$ as the embedded feature that is extracted from predicted probabilities.
We firstly consider $\mathbf{p} = (p_{1}, p_{2},\ldots,p_{M})$ as \textit{probability} vector, where $M$ is the number of sound events defined in AudioSet dataset~\cite{audioset} and $p_{i}$ is the total sum of predicted probabilities of the $i$th sound event detected in each programme.
The embedding $\mathbf{mean{-}prob{-}k} = (\bar{p}_{1}, \bar{p}_{2},\ldots,\bar{p}_{M})$ is then computed by $\bar{p}_{i} = \frac{p_{i}}{\sum_{i=1}^{M}p_{i}}$.
%
%
%
\subsection{Candidate Back-end Classifiers}

\begin{table}[t]
    \caption{Different Back-end Classifier Evaluated.} 
        	\vspace{-0.2cm}
    \centering
    \scalebox{0.95}{
    \begin{tabular}{l c} 
        \hline 
            \textbf{Classification Models}   &  \textbf{Setting parameters}  \\
        \hline 
        LR (baseline) & -\\
        \hline     
        SVM  & C=1.0\\
        &  Kernel=`RBF' \\
                \hline                   
        DT & Max Depth of Tree = 20\\
                \hline            
        RF & Max Depth of Tree = 20\\
        & Number of Trees = 100\\
                \hline            
        MLP  &  Output Dimension\\        
         FC - ReLU - Dropout (20\%)&  $2048$       \\
         FC - ReLU - Dropout (30\%)&  $4096$       \\
         FC - ReLU - Dropout (40\%)&  $4096$       \\
         FC - ReLU - Dropout (50\%)&  $1024$       \\
         FC - Softmax  &  $9$       \\
       \hline 
    \end{tabular}
    }
    \label{table:post_train} 
\end{table}
\begin{table}[t]
    \caption{BBC television programme dataset} 
    \centering
   \scalebox{0.95}{
    \begin{tabular}{l c } 
        \hline 
	    Genres  & Number of programmes\\
        \hline 
        Children's                  & 698            \\
        Drama                      & 695         \\
        Factual                    & 692               \\
        Music            & 670             \\
        News                     & 700           \\
        Weather           & 654               \\
        Sport            & 663           \\
        Comedy        & 690             \\
        Entertainment           & 698                \\
        \hline 
        Total programmes              &  6,160         \\
        \hline 
    \end{tabular}
    }
    \label{table:bbc_data} 
\end{table}
\begin{table*}[t]
    \caption{Performance of back-end classification models (Linear Regression(LR),  Decision  Tree  (DT),  Support  Vector  Machine  (SVM),  Random  Forest(FR),  and  Multilayer  Perceptron  (MLP)) with different statistics (average Accuracy over 14 folds) .} 
        	\vspace{-0.2cm}
    \centering
    \scalebox{0.85}{
    \begin{tabular}{c  c c  c c  c c  c c  c c} 
        \hline 
            \textbf{Back-end} &\textbf{mean-num-1}  & \textbf{mean-prob-1}  &\textbf{mean-num-4}  & \textbf{mean-prob-4} &  \textbf{mean-num-6}  & \textbf{mean-prob-6} &  \textbf{mean-num-8}  & \textbf{mean-prob-8} &  \textbf{mean-num-10}  & \textbf{ mean-prob-10}  \\
                                                  \textbf{Classifiers}  &  \textbf{(\%)}  & \textbf{(\%)}  & \textbf{(\%)} &  \textbf{(\%)}  & \textbf{(\%)}  & \textbf{(\%)} &  \textbf{(\%)}  & \textbf{(\%)}  &  \textbf{(\%)}  & \textbf{(\%)} \\

        \hline 
	    LR     &56.7(\textbf{baseline})          &56.1             &85.9           &83.9           &88.0          &85.8           &88.6         &86.9           & 89.1          &87.4  \\
        \hline                                                       
        SVM    &35.1          &34.2             &84.9           &73.9           &87.2          &75.7           &87.5         &76.8           & 88.3          &77.2 \\
        \hline                                                       
        DT     &61.5          &60.2             &79.8           &78.9           &79.0          &79.5           &80.2         & 79.3          & 80.5          &79.4  \\
        \hline                                                       
        RF     &\textbf{69.4} &\textbf{68.5}    &90.9           &90.5           &91.4          &90.8           &91.4         &91.1           &91.5           &91.1 \\
        \hline                                              
        MLP    &62.1          &60.9            &\textbf{92.6}  &\textbf{91.6}  &\textbf{93.2} &\textbf{92.1}  &\textbf{93.5} &\textbf{92.2} & \textbf{93.7} & \textbf{92.4} \\                 
       \hline 
    \end{tabular}
    }
    \label{table:res_fea} 
\end{table*}
\begin{table}[th]
    \caption{Performance of back-end classification models with the combined feature (average Accuracy over 14 folds)} 
        	\vspace{-0.2cm}
    \centering
    \scalebox{0.95}{
    \begin{tabular}{c c c c} 
        \hline 
            \textbf{Back-end}   &  \textbf{mean-num-10}  & \textbf{mean-prob-10}  & \textbf{combined-feature}  \\
                               \textbf{Classifiers}         &  \textbf{(\%)}  & \textbf{(\%)}  & \textbf{(\%)}  \\
        \hline 
        LR & 89.1  & 87.4 & 89.7\\
        \hline            
        SVM & 88.3 & 77.2 & 83.6\\
        \hline            
        DT   & 80.5 & 79.4  & 80.4\\
                \hline            
        RF & 91.5 & 91.1 & 91.8 \\
                \hline            
        MLP  & \textbf{93.7} & \textbf{92.4} & \textbf{93.6}\\                 
       \hline 
    \end{tabular}
    }
    \label{table:combine_feature} 
\end{table}

In addition to Linear Regression (LR) used for classification in the baseline system, we further evaluate different back-end classification models.
In particular, we apply both traditional machine learning models (Support Vector Machine (SVM), Decision Tree (DT), Random Forest (RF)) and a deep learning based model using Multilayer Perceptron (MLP) network architecture.
The back-end classification models evaluated are configured as shown in Table \ref{table:post_train}.

\section{Evaluation Method}
\subsection{Dataset}
\textbf{AudioSet}: This is a large-scale dataset released by Google~\cite{audioset}, in which there are around 2,084,320 10-second audio clips. This dataset contains a total of 527 sound event classes (including both natural sounds and human speeches) with annotation. Each 10-second audio clip may contain more than one type of sound events and there is no information of onset and offset for a certain sound event (i.e. weakly labelled dataset of sound events). This dataset is used for training the pre-trained CNN network as mentioned in our previous work~\cite{kong_pretrain}.

\textbf{BBC programmes}: The dataset is collected by selecting programmes from eight different BBC TV channels (BBC One, BBC Two, BBC Four, CBBC, CBeebies, BBC News 24, BBC Parliament, and BBC World News)~\cite{redux}. 
These programmes are from the nine main BBC genre categories~\cite{genres} (\textit{Children's, Drama, Factual, Music, Sport, Weather, Comedy, Entertainment} and \textit{News}).
Approximately 25 programmes per genre are collected for each month in the years 2019 and 2020, creating a dataset of 6,160 BBC programmes as shown in Table \ref{table:bbc_data}. The audio is MP3-encoded at 128kbps joint stereo.
To evaluate, we separate this dataset into 14-fold cross validation and report the final classification accuracy as an average over 14 folds. 

\subsection{Experimental setting}
We construct the pre-trained CNN network by using Pytorch framework~\cite{kong_pretrain}. 
We use the following cross-entropy loss function during training of this network as
\begin{equation}
    \label{eq:loss_func}
    Loss_{EN}(\Theta) = -\frac{1}{N}\sum_{n=1}^{N}\mathbf{y_n} \log \left\{\mathbf{\hat{y}_{n}}(\Theta) \right\} 
\end{equation}
defined over all parameters \(\Theta\), and $N$ is the number of training clips.  $\mathbf{y_{n}}$ and $\mathbf{\hat{y}_{n}}$  denote ground truth and predicted output.
The training is carried out for 100 epochs using Adam ~\cite{Adam} for optimization.
Regarding the back-end classification models, we use Scikit-learn toolbox~\cite{scikit-learn} to implement traditional machine learning models such as Linear Regression (LR), Support Vector Machine (SVM), Decision Tree (DT), and Random Forest (RF). 
For Multilayer Perceptron (MLP), we also construct the network by the Pytorch framework. The loss function used for training the MLP based network is also cross-entropy (\ref{eq:loss_func}). 
To enforce the MLP based network, we apply \textit{mixup} data augmentation~\cite{mixup1, mixup2} to audio embedded input features.

Regarding the evaluation metric used in this paper, if $C$ is considered as the number of audio recordings of programmes which are correctly predicted, and the total number of audio recordings is $T$, the classification accuracy (Accuracy (\%)) is the \% ratio of $C$ to $T$.
%
%
%
%

\section{Results}
\begin{figure}[th]
    \centering
    \includegraphics[width =1.0\linewidth]{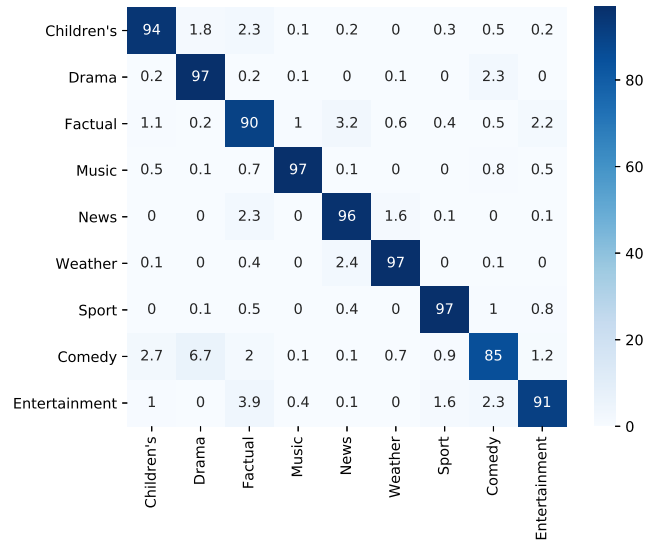}
    	\vspace{-0.5cm}
	\caption{Confusion matrix result (\%) with $\mathbf{mean{-}num{-}10}$ embedding and back-end MLP (average Accuracy over 14 folds) }
    \label{fig:R1}
\end{figure}
Experimental results over both embedded features and all back-end classification models are presented in Table \ref{table:res_fea}. 
As shown in Table \ref{table:res_fea}, when the number of detected sound events used for classification increases, the accuracy is improved over all back-end classification models.
It can be concluded that multiple-sound-event tagging is beneficial to extract embedding features rather than single-sound-event tagging.
Comparing between the two types of embedding features, the sound event based embeddings perform better than predicted probability based embeddings over all back-end classifiers.
Regarding back-end classification models evaluated, Random Forest and MLP based network outperform Linear Regression, Decision Tree and Support Vector Machine.
The best score of 93.7\% is obtained from MLP based network with $\mathbf{mean{-}num{-}10}$ embedding, significantly improving the baseline of $\mathbf{mean{-}num{-}1}$ embedding and back-end Linear Regression by approximately 37.0\% .

The confusion matrix in Fig. \ref{fig:R1} shows an average over 14 folds with $\mathbf{mean{-}num{-}10}$ embedding and back-end MLP model. 
It can be seen that wrong inference occurs among related programmes such as \textit{Entertainment} and \textit{Comedy}, or programmes of \textit{News} and \textit{Weather}.
Meanwhile, \textit{Music, Sport, Weather} and \textit{Drama} achieve the best performance of aprroximately 97.0\% among genres.
\section{Further Experiments}
We also conduct further experiments to evaluate whether combination of both sound event based embedded feature and predicted probability based embedded feature, referred to as  $\mathbf{combined{-}feature}$, can help to improve the performance.
To this end, two embeddings are concatenated before feeding into the back-end classification models. 
As obtained results in Table \ref{table:combine_feature}, $\mathbf{combined{-}feature}$ (a concatenation of $\mathbf{mean{-}prob{-}10}$ and $\mathbf{mean{-}num{-}10}$) helps Linear Regression and Random Forest improve the performance, but not effective for the other models.
\begin{figure}[t]
    \centering
    \includegraphics[width =1.0\linewidth]{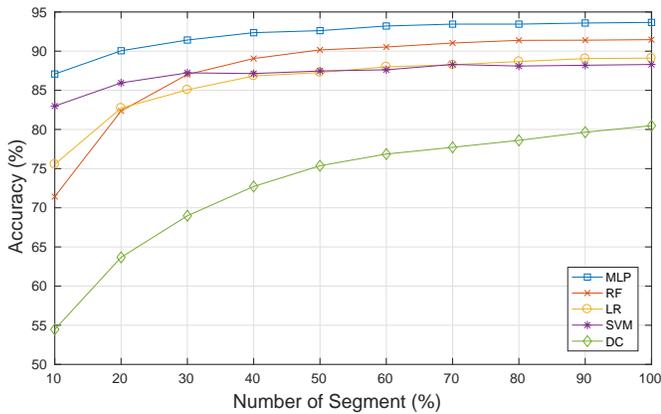}
    	\vspace{-0.5cm}
	\caption{Performance of all back-end classification models (Linear Regression (LR), Decision Tree (DT), Support Vector Machine (SVM), Random Forest (FR), and Multilayer Perceptron (MLP)) with reduced number of input segments and $\mathbf{mean{-}num{-}10}$ embedding (average of 14 folds)}
    \label{fig:line_pc}
\end{figure}

As the duration of BBC programmes can be long, we evaluate whether a programme can be effectively detected with a reduced number of input segments, thus help to reduce the cost of inference process.
In particular, 10\% to 100\% of the input segments are randomly selected from each programme for evaluation.
We use $\mathbf{mean{-}num{-}10}$ embedding and all back-end classification models.
As shown in Fig. \ref{fig:line_pc}, if 60\% of segments or more are used, almost post-trained models' performance apart from Linear Regression is stable.
Notably, MLP classifier achieves an average classification accuracy of 93.0\% that potentially reduces 40\% of time for the inference process.

\section{Conclusion}
We have explored a deep learning based framework for classifying BBC television programmes into nine genres that match the BBC genre categories.
Our framework, which uses a log-mel spectrogram representation, a pre-trained CNN architecture for extracting embedded features, and a back-end Multilayer Perceptron classifier, achieved an average classification accuracy of 93.7\% over 14-fold cross validation.
In further work, we will evaluate whether the audio-based embedded features can be used to measure the similarity between BBC television programmes for the purpose of recommendations.

\section*{Acknowledgement}
This work was funded by an EPSRC Impact Acceleration Account project EP/R511791/1, and carried out jointly by the University of Surrey and British Broadcasting Corporation (BBC).


\small

\bibliographystyle{IEEEtran}
\bibliography{refs}

\end{document}